\documentclass[sigconf]{acmart}

\AtBeginDocument{%
  \providecommand\BibTeX{{%
    \normalfont B\kern-0.5em{\scshape i\kern-0.25em b}\kern-0.8em\TeX}}}

\usepackage{indentfirst}
\usepackage{listings}
\usepackage{subfigure}
\usepackage{caption}
\usepackage{xcolor}

\begin{document}

\title{Supporting CUDA for an extended RISC-V GPU architecture }

\author{Ruobing Han}
\email{hanruobing@gatech.edu}
\affiliation{%
  \institution{Georgia Institute of Technology}
  \country{USA}
}

\author{Blaise Tine}
\email{blaisetine@gatech.edu}
\affiliation{%
  \institution{Georgia Institute of Technology}
  \country{USA}
}

\author{Jaewon Lee}
\email{jaewon.lee@gatech.edu}
\affiliation{%
  \institution{Georgia Institute of Technology}
  \country{USA}
}

\author{Jaewoong Sim}
\email{jaewoong@snu.ac.kr}
\affiliation{%
  \institution{Seoul National University}
  \country{Korea}
}

\author{Hyesoon Kim}
\email{hyesoon@cc.gatech.edu}
\affiliation{%
  \institution{Georgia Institute of Technology}
  \country{USA}
}

\begin{abstract}
    With the rapid development of scientific computation, more and more researchers and developers are committed to implementing various workloads/operations on different devices. Among all these devices, NVIDIA GPU is the most popular choice due to its comprehensive documentation and excellent development tools. As a result, there are abundant resources for hand-writing high-performance CUDA codes. However, CUDA is mainly supported by only commercial products and there has been no support for open-source H/W platforms. RISC-V is the most popular choice for hardware ISA, thanks to its elegant design and open-source license. In this project, we aim to utilize these existing CUDA codes with RISC-V devices. More specifically, we design and implement a pipeline that can execute CUDA source code on an RISC-V GPU architecture. We have succeeded in executing CUDA kernels with several important features, like multi-thread and atomic instructions, on an RISC-V GPU architecture.

\end{abstract}


\maketitle

\section{Introduction}

RISC-V is the most popular choice for researchers in the academic community and engineers in hardware companies. The most important reason is its open-source spirit. These open-source licenses encourage many researchers to devote themselves to the development of a mature ecology for RISC-V, and thus, in turn, more and more people are willing to join the community, as there are existing fancy codes, hardware designs, and so on.

In the RISC-V ecology, the software support is the bottleneck for the blooming of the RISC-V community. Although OpenCL is an open platform for heterogeneous computing, due to the stability and software tool chain support, CUDA has been used widely. Unfortunately, CUDA source code can only be compiled and then executed on NVIDIA's devices, which is a major obstacle to using RISC-V for a wide range of applications, especially high-performance computing and machine learning workloads.

One way to solve this dilemma is to use code migration\cite{kuznetsov2019porting,kontogiannis2010code,nguyen2016mapping}. Instead of using the default method to compile CUDA source code with NVIDIA's compiler, some researchers try to parse and modify the source code to other high-level languages; more detail is shown in Sec. \ref{sec:program_migration}. However, because these methods highly rely on the high similarity between CUDA and the target high-level languages, they are not general solutions. Another solution is to build a compiler that directly compiles high-level CUDA language into a low-level RISC-V binary file. To the best of our knowledge, although there are translators that support generating RISC-V, none of them can handle CUDA source code.

\begin{figure*}[htbp]
    \centering
    \includegraphics[width=160mm]{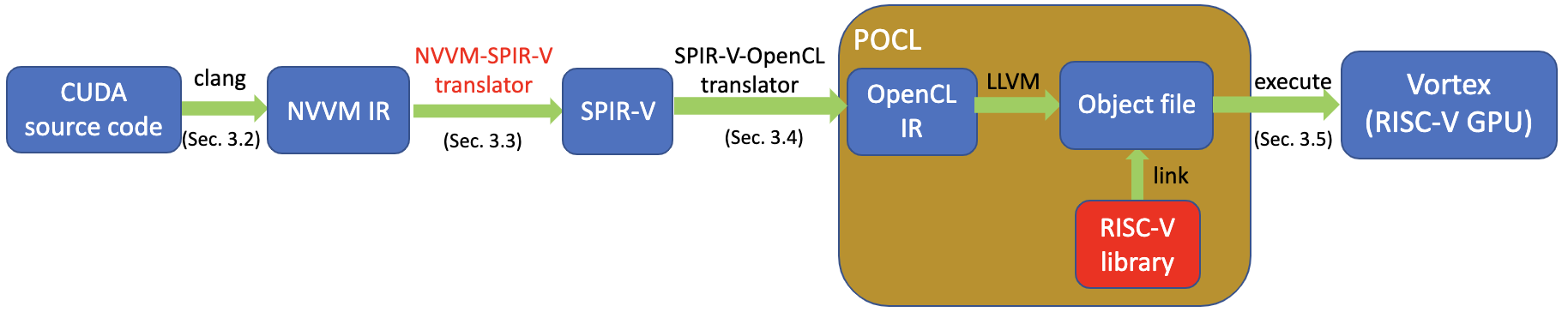}
    \caption{Overview of pipeline. The red part (NVVM-SPIR-V translator and RISC-V library) is developed in this work.}
    \label{fig:pipeline}
\end{figure*}

Thus, in this project we propose and build a pipeline to support an end-to-end CUDA migration: the pipeline accepts CUDA source codes as input and executes them on an extended RISC-V GPU architecture. Our pipeline consists of several steps: translates CUDA source code into NVVM IR\cite{grover2012compiling}, converts NVVM IR into SPIR-V IR \cite{kessenich2018spir}, forwards SPIR-V IR into POCL\cite{jaaskelainen2015pocl} to get RISC-V binary file, and finally executes the binary file on an extended RISC-V GPU architecture. We choose to use an intermediate representation (SPIR-V) for two reasons 1) RISC-V is still in development and has a lot of extensions, so we should not directly convert CUDA into RISC-V, as it will make supporting new features in the future difficult for our pipeline; 2) we want to make our pipeline more general so that we can support CUDA as front-end and RISC-V as back-end. Our pipeline is represented by Fig. \ref{fig:pipeline}.

In conclusion, the main contributions of our paper include the following:
\begin{itemize}
    \item propose and implement a pipeline for executing CUDA source code on RISC-V GPU;
    \item build a translator support translating from NVVM to SPIR-V\footnote{https://github.com/gthparch/NVPTX-SPIRV-Translator};
    \item pipeline that is easy to maintain and further support other front-end languages and back-end devices.
    \item pipeline that is lightweight, which can be executed without NVIDIA GPUs
    \item extend existing POCL to support RISC-V back-end
\end{itemize}

The rest of this paper is organized as follows. Section \ref{sec:releated_work} provides a survey that describes various attempts to migrate CUDA source code and introduces the current Available frames used in our pipeline. Section \ref{sec:overview_pipeline} uses a simple example to go through the entire pipeline to give a brief introduction of each phase in the pipeline.
Section \ref{sec:experiment} records how to support several important features in CUDA.
Finally, our concluding thoughts are in Section \ref{sec:conclusion}.

\section{Background and related work\label{sec:releated_work}}

\subsection{Program migration\label{sec:program_migration}}
Many existing works\cite{perkins2017cuda,babej2020hipcl} aim to migrate from CUDA to other representations for different back-end devices.

HIPIFY\footnote{\url{https://github.com/ROCm-Developer-Tools/HIPIFY}} is a tool to translate CUDA source code into portable HIP C++, another high-level language proposed for AMD devices. HIPIFY has two methods to translate CUDA to HIP C++: hipify-perl and hipify-clang. Hipify-perl is quite straightforward; it heavily uses regular expressions to translate CUDA keywords to corresponding HIP C++ keywords. Instead of word-to-word replacement, Hipify-clang uses a more complex method: it adds an adaptor in Clang. It uses Clang's syntax analysis, converts CUDA source code into an abstract syntax tree, uses transformation matchers to convert this syntax tree, and finally generates HIP C++. Both methods can only be used for HIP C++ or other source languages whose grammars are highly closed to CUDA. HIPIFY is a translator from a high-level language to another high-level language, as most of its workload is done at the lexical level.
To support executing CUDA on AMD devices, users need to first use HIPIFY to convert CUDA into HIP C++ and then compile the generated code with the HIP C++ compiler. We cannot use the same method to migrate CUDA on RISC-V devices, as there is not a corresponding high-level language for RISC-V.

SYCL\cite{silveira2000dpc++,keryell2015khronos}, proposed by the Khronos group, is another project that focuses on deploying source kernels on different devices. It is a high-level programming model that aims to improve programming productivity on various hardware accelerators. It can be regarded as a series of libraries of C++. These libraries provide APIs needed to write programs that can be deployed and executed in various back-end devices without modifying the code. This is much like OpenCL, which also uses C/C++ for high-level users, and generates host/device programs and executes them automatically for different back-end devices. Compared with OpenCL, SYCL is at a higher level; it provides primitives to support users to implement programs for devices directly, instead of regarding the code as a string and directly forwarding it to back-end devices' drivers. Thus, SYCL is highly portable thanks to its high-level abstraction.
However, SYCL cannot solve our dilemma, as it does not support CUDA. SYCL can not migrate the existing CUDA source code to execute it with RISC-V. Instead, it requires users to re-implement it with APIs provided by SYCL.

\subsection{Intermediate Representation}
\begin{figure*}[htbp]
    \centering
    \subfigure[NVVM]{
    \begin{minipage}[t]{0.3\linewidth}
    \centering
        \includegraphics[width=1.0\linewidth]{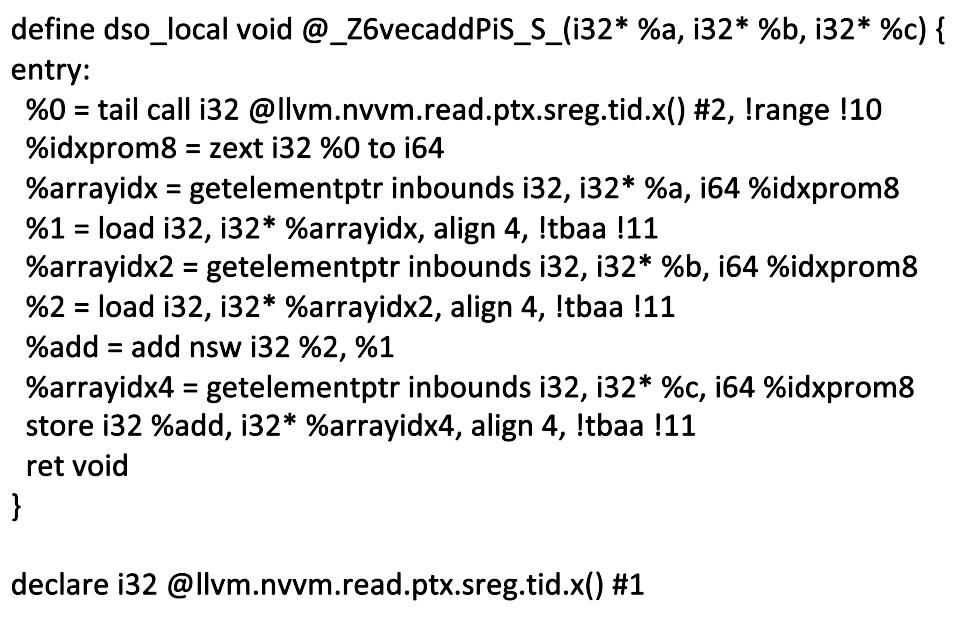}
    \end{minipage}%
    }%
    \subfigure[SPIR-V]{
    \begin{minipage}[t]{0.3\linewidth}
    \centering
        \includegraphics[width=1.0\linewidth]{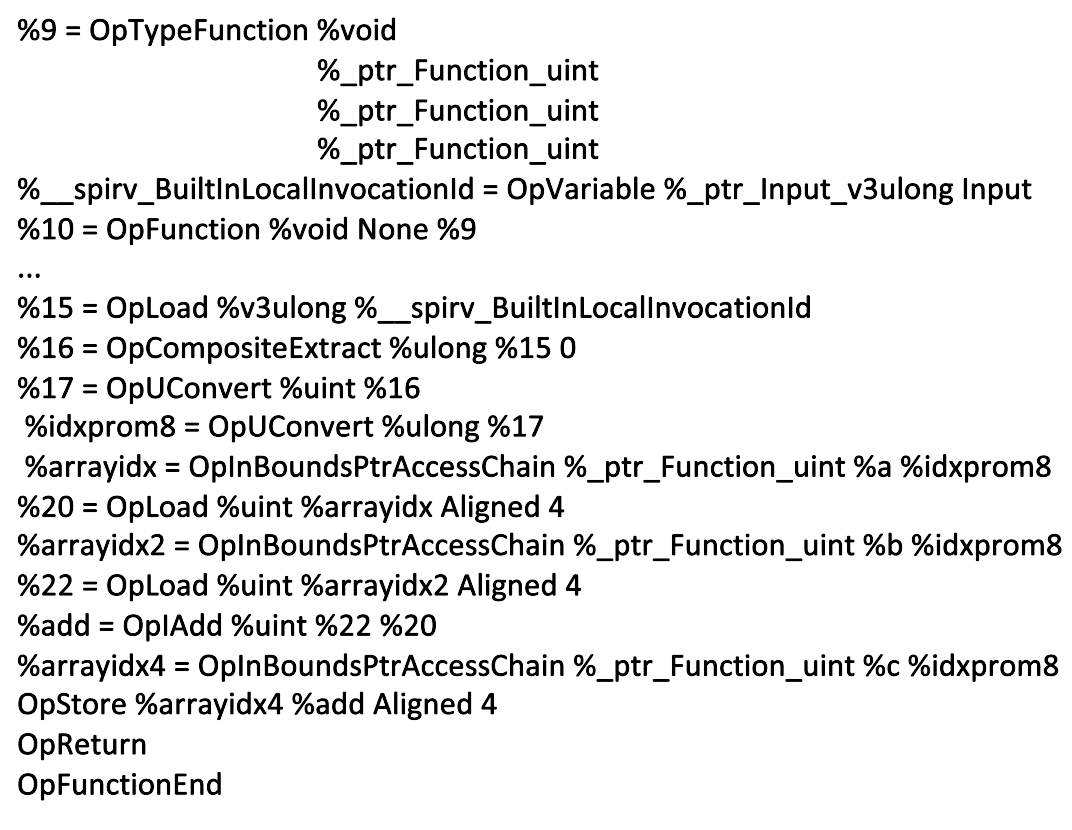}
    \end{minipage}%
    }%
    \subfigure[OpenCL IR]{
    \begin{minipage}[t]{0.3\linewidth}
    \centering
        \includegraphics[width=1.0\linewidth]{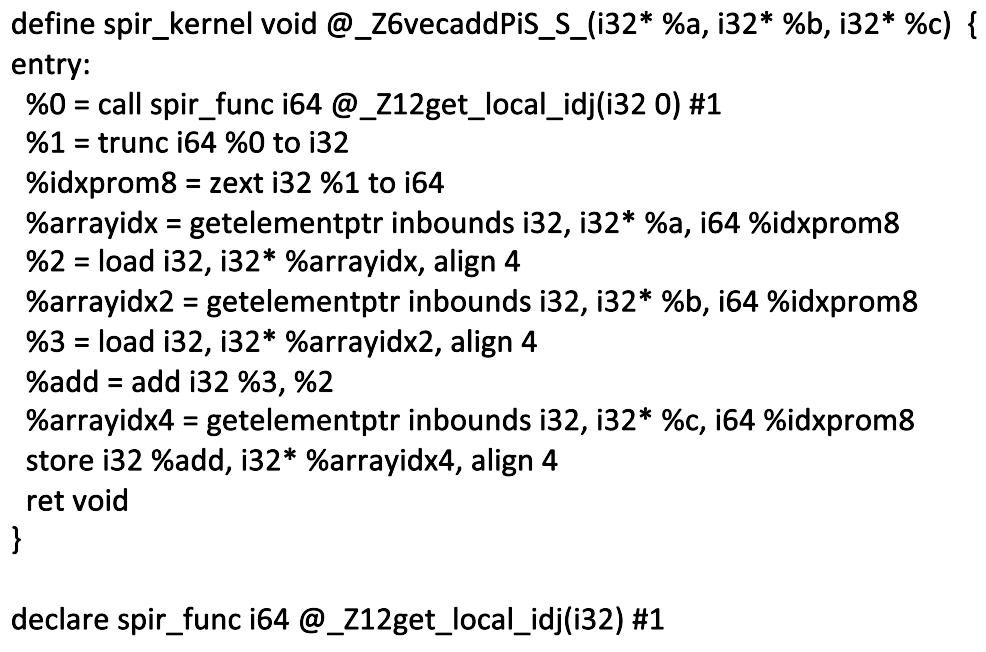}
    \end{minipage}
    }%
    \centering
    \caption{Implementation of vecadd in NVVM/SPIR-V/OpenCL IR}
    \label{fig:different_IR}
    \end{figure*}

In modern compiler design, a compiler can always generate object files for different back-end devices with various input languages. To support multiple source languages with multiple back-end devices, compilers first compiles different source languages to a standard intermediate representation (IR) and then emits this IR into different binary files according to our target back-end devices.
Our pipeline has three involved IRs: NVVM IR, SPIR-V, and OpenCL IR. In Fig. \ref{fig:different_IR}, we show how these three IRs represent a vecadd example. IRs in Fig. \ref{fig:different_IR}(a) and (c) have similar format, most of instructions are same except some call instructions. These is due to the difference between NVVM and OpenCL IR is only for built-in functions: they have different built-in function( $llvm.nvvm.read.ptx.srge.tid.x$ and $get_local_id$) for a same primitive (get the x-dim thread index). SPIR-V IR is total different with NVVM and OpenCL IR, not only for different built-in functions, but also different instructions for load/store/binary op etc.

\subsubsection{NVVM}
NVVM IR\cite{grover2012compiling}, proposed by NVIDIA, is a compiler IR used to represent GPU compute kernels. In the general case, to execute CUDA source code on NVIDIA GPUs, we need to call nvcc to compile CUDA into NVVM IR, and the GPU driver will further translate the NVVM IR into a binary code that can be run on the processing cores of NVIDIA GPUs. NVVM IR is highly compatible with LLVM\cite{lattner2004llvm}; it supports a subset of LLVM IR along with a defined set of the build-in functions used to represent GPU programming concepts.

\subsubsection{SPIR-V}
SPIR-V \cite{kessenich2018spir} is the fifth version of SPIR.
SPIR (Standard Portable Intermediate Representation) is an intermediate representation (IR) used for expressing parallel computation and GPU-based graphics.
SPIR is a general IR that can allow high-level users to use various front-end's programming languages. SPIR-V can be used as an isolation layer that departs the front-ends programming language (e.g. MLIR, OpenCL, OpenGL) and low-level compute architecture. With SPIR, we first compile the source code into SPIR IRs and forward these IRs to devices. Thus, we can eliminate the need for high-level language front-end compilers in device drivers. This will also relieve the kernel launch time, as we do not deal with complex and abstract high-level languages, but rather the hardware-friendly SPIV IR. Besides, with SPIR as an isolation layer, high-level users do not care about low-level hardware; they can choose to use any programming language they like without worry about the reliability and portability of their programs. For hardware researchers, they can only focus on the optimization for the SPIR without worry about front-end language.

\subsubsection{OpenCL IR}
OpenCL (Open Computing Language) \cite{munshi2009opencl} is a framework for writing programs that execute across heterogeneous platforms. It supports several back-end devices: CPUs, GPUs, FPGAs, and so on.
Although OpenCL can support using the native language (CUDA, HIP C++ etc.) to implement the kernel for each back-end device, it also provides OpenCL C programming language, a series of high-level APIs used for high-performance computing. This language provides a rich set of built-in functions for scalar and vector operations. These functions support scalar and vector argument types and can be used to implement high-performance programs for different back-end devices. When we refer to LLVM-IR, we generally refer to LLVM bc format of LLVM-IR.

\subsection{Current Available Frames}

\subsubsection{OpenCL-SPIR-V translator\label{sec:LLVM-SPIRV}}
The Khronos Group developed a Bi-Directional translator\footnote{\url{https://github.com/KhronosGroup/SPIRV-LLVM-Translator}} that supports converting between OpenCL IR and SPIR-V files.
Although it generates SPIR-V, which can further be deployed and executed with RISC-V back-end devices, it only accepts OpenCL IR as input.
In our project, we add some extensions based on the original translator to support its handling NVVM.
As described in Sec. \ref{sec:releated_work}, NVVM is a subset of LLVM IR along with a defined set of the built-in functions. The translator cannot handle these NVVM-specific built-in functions. For example, in Fig. \ref{fig:different_IR}, (c) is an OpenCL IR and can be translated by the translator. However, (a) is an NVVM IR and has an NVVM-specific built-in function,  $llvm.nvvm.read.ptx.srge.tid.x()$. When we pass this NVVM to the translator, it will raise an error, as it does not recognize this built-in function, but only recognize $get\_local\_id$ in OpenCL IR. 

\subsubsection{POCL}
POCL (Portable Computing Language) \cite{jaaskelainen2015pocl} is being developed as an efficient implementation of the OpenCL standard. POCL is a framework that accepts SPIR-V binary files for input. POCL will first convert SPIR-V into OpenCL IR, using the translator mentioned above, links the converted OpenCL IR with the runtime library, and finally emits the object file. POCL can support several back-end devices, like x86, CUDA, MIPS, etc. POCL is highly extensible; to support a new back-end device, researchers only need to provide a runtime API library for OpenCL IR, and this library will be used to link when emitting object files. Although original POCL does not support RISC-V, other researchers \cite{elsabbagh2020vortex} have supported RISC-V on POCL and executed these file on several back-end devices. In our experiment, we use this version of POCL.\footnote{\url{https://github.com/vortexgpgpu/pocl}}

\subsubsection{Vortex}
Vortex\footnote{\url{https://vortex.cc.gatech.edu/}}\cite{elsabbagh2020vortex} is an open-source RISC-V-based GPGPU processor. Vortex implements a SIMT architecture with a minimal ISA extension to RISC-V ({\it tmc}: activate  threads, {\it wspan}: spawn a waive-front (or warp), {\it split/join}: divergent branch handling instructions, {\it bar} : stall waive-front ). Currently, Vortex can execute OpenCL IR through POCL runtime systems. Thus, we can directly execute our generated object file on Vortex. We also use Vortx's RISC-V library in the linkage phase in POCL.

\section{Overview of the pipeline\label{sec:overview_pipeline}}

In this section, we show how to execute a simple vector add CUDA source code (Code. \ref{code:vecadd_cuda}) on an RISC-V GPU. 

\subsection{Input CUDA source code\label{sec:source_code}}
CUDA can be regarded as an extension of standard C++. The only difference is that CUDA has some other extra features, like definitions of memory hierarchy and some unique built-in functions. As it's quite easy to execute C++ on RISC-V back-end devices, the only part we need to handle is the extra part specific for CUDA. For example, in our example, only two CUDA specific features do not belong to standard C++: $\_\_global\_\_$ and $threadIdx.x$. $\_\_global\_\_$ is a CUDA C keyword that says the function should be called from the host. $threadIdx.x$ can be regarded as a built-in function that records the work-item's local index in x-dim.

\begin{lstlisting}[caption={VectorAdd CUDA source code},label={code:vecadd_cuda},language=C]
__global__ void vecadd (int *a, int *b, int *c) {
  int gid = threadIdx.x;
  c[gid] = a[gid] + b[gid];
}
\end{lstlisting}

\subsection{Compile to NVVM}
Our pipeline will directly use the CUDA toolkit to compile CUDA source code into NVVM IRs. Thus, we can get the following NVVM IR (Code. \ref{code:vecadd_nvptx})(For clarity, we remove some unimportant content):

\begin{lstlisting}[caption={NVVM VectorAdd IR},label={code:vecadd_nvptx},language=C]
target triple = "nvptx64-nvidia-cuda"

define dso_local void @vecadd(
    i32* nocapture readonly %a, 
    i32* nocapture readonly %b, 
    i32* nocapture %c) {
entry:
  ; see Fig. 2(a) for deatil code
}
declare i32 @llvm.nvvm.read.ptx.sreg.tid.x()

!nvvm.annotations = !{!3}
!3 = !{void (i32*, i32*, i32*) *@vecadd, 
        !"kernel", i32 1}

\end{lstlisting}

At the beginning, the IR has a meta-data variable $target$ $triple$. This meta-data records the target back-end device, which is used for further code generation to emit object files. For NVVM, it will always have value $"nvptx64-nvidia-cuda"$.

$nvvm.annotations$ is another important meta-data. It records the kernel function of this program. As mentioned in Sec. \ref{sec:source_code}, CUDA has a special keyword, $\_\_global\_\_$, to mark a function as a kernel function. As we cannot use this keyword in NVVM, we have to use an extra meta-data to record this information.

This NVVM IR has a function $llvm.nvvm.read.ptx.srge.tid.x$, which is declared but not defined. This is a built-in function that will be linked with CUDA libraries. As these libraries can only be used for NVIDIA GPUs, we have to replace these built-in functions in the next phase.
In Table. \ref{table:built-in}, we show several built-in functions that are widely used in high-performance computing programs.
\begin{table}[]
\begin{tabular}{|c|c|}
\hline
function name                 & detail                             \\ \hline
llvm.nvvm.read.ptx.sreg.ctaid & get the block index                \\ \hline
llvm.nvvm.read.ptx.sreg.ntid  & get the block dimension            \\ \hline
llvm.nvvm.read.ptx.sreg.tid   & get the thread index               \\ \hline
llvm.nvvm.barrie              & synchronize threads within a block \\ \hline
llvm.sqrt                     & calculte the square root           \\ \hline
llvm.fabs                     & calculate the absoulte value       \\ \hline
llvm.nvvm.d2i                 & narrowing convertions              \\ \hline
llvm.fma                      & fused multiply–add                 \\ \hline
\end{tabular}
\caption{Built-in functions that are widely used in high-performance computing programs.}
\label{table:built-in}
\end{table}

\subsection{Translate NVVM to SPIR-V}
    \begin{figure}[htbp]
    \centering
    \includegraphics[width=70mm]{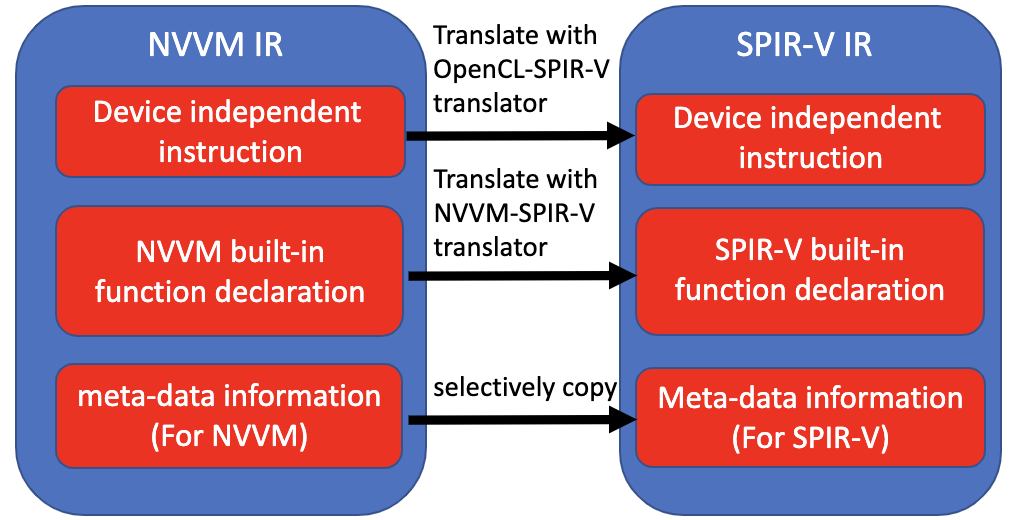}
    \caption{An overview of our translator.}
    \label{fig:translator_overview}
    \end{figure} 

In Fig. \ref{fig:translator_overview}, we show an overview of our translator. The input NVVM IR comprises three components: metadata (e.g. device version, function property), NVVM built-in function declaration (e.g. functions for getting thread index), and device-independent instruction.

\subsubsection{Meta-data\label{label:meta-data}}
Meta-data is used to record information like function name, data layout, and address length. Most meta-data are independent with back-end devices; thus we can directly copy them. Thus, in our vecadd example, we remove all meta-data except these back-end related meta-data. For a target triple, we have to modify it to the corresponding SPIR-V triple. As for some CUDA-specific meta-data, we have to either directly drop it or translate it to corresponding SPIR-V meta-data. 
In NVVM, $nvvm.annotations$ is used to denote the kernel function for the whole program, which is marked with $\_\_global\_\_$ in CUDA source code. However, for SPIR-V, it uses a different method to record the kernel functions: these functions that have meta-data $kernel\_arg\_type$. Thus, for each given NVVM, we have to record the functions recorded by $nvvm.annotations$, remove this meta-data, and add a new meta-data $kernel\_arg\_type$ for these functions.
Finally, some meta-data are not needed in NVVM due to some CUDA's hypotheses. However, these hypotheses do not exist in SPIR-V. Thus, we have to add these meta-data explicitly. For example, in CUDA, when the input variables are pointer, they should always point to the global memory. So there is no meta-data to represent the memory hierarchy for each input pointer. While SPIR-V does not have this limitation, we have to add an extra meta-data $kernel\_arg\_addr\_space$ to record this information, to tell the compile and back-end devices that these pointers are all point to the address in global memory.

\subsubsection{Built-in function}
Built-in functions are functions that are special for the given back-end devices or framework. They are the key for each device, as they can be regarded as primitives of frameworks. 
Although for the Single-Instruction-Multiple-Data protocol, all devices have the same primitives: for getting threads' indexes, getting blocks' indexes, getting the length of each dimension, and so on. They have different built-in functions to implement these primitives.

For example, in NVVM, programs directly call a built-in function $llvm.nvvm.read.ptx.srge.tid.x$ to get the global thread index in x-dimension for a thread. SPIR-V uses a different method; in SPIR-V, a variable $GlobalInvocationId$ records a thread's index in different dimensions. To get the index, we have to first load this variable from memory and then extract the index for a special dimension.

\subsubsection{Device independent instruction}
\begin{figure*}[htbp]
    \centering
    \includegraphics[width=170mm]{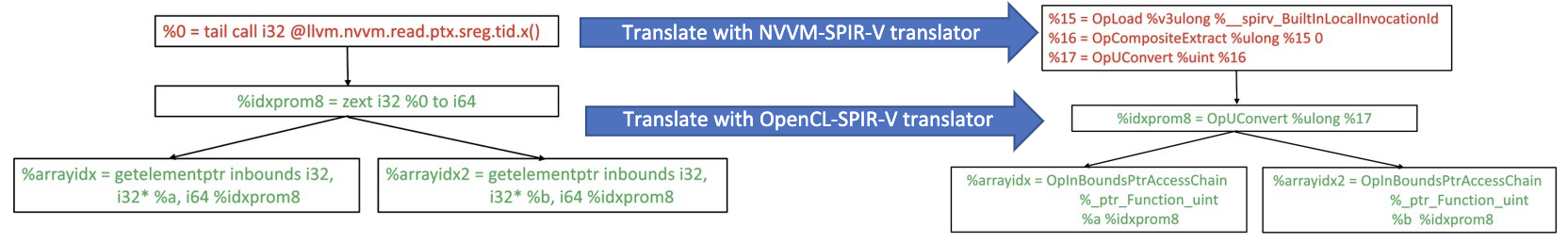}
    \caption{For NVVM specific built-in function, we have to handle the corresponding instructions (red text) with NVVM-SPIR-V translator. While for device independent instructions (green text), we can directly translate them with existing OpenCL-SPIR-V translator}
    \label{fig:device_independ_inst}
    \end{figure*}

In the above two sections, we modify contents that are different between NVVM and OpenCL IR, which cannot be translated by the existing OpenCL-SPIR-V translator. 
However, usually only a small part of a program belongs to these two classes, and most instructions are device independent. For these instructions, we can directly use the existing OpenCL-SPIR-V translator. For example, in Fig. \ref{fig:device_independ_inst}, we have four instructions in NVVM IR. Only the first instruction (shown with red text) is NVVM-specific and has to be translated with the NVVM-SPIR-V translator. The rest instructions (shown with green text) are device independent can be translated with the existing OpenCL-SPIR-V translator.

\begin{figure}[htbp]
    \centering
    \includegraphics[width=70mm]{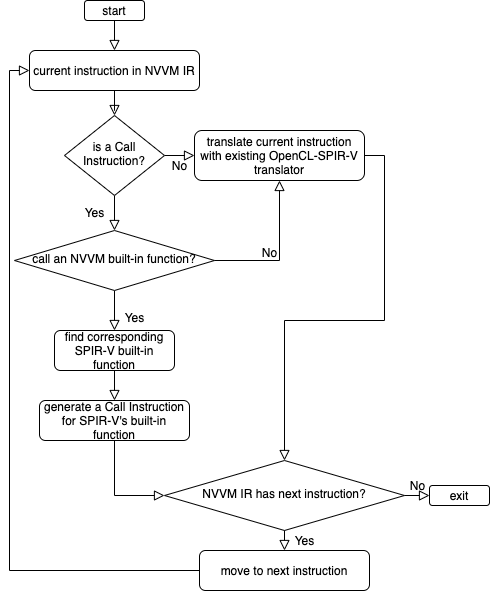}
    \caption{Our translators will reuse the OpenCL-SPIR-V translator, except when a instruction calls NVVM built-in function.}
    \label{fig:translator_diagram}
    \end{figure}

Separately handling device-dependent/independent instructions can avoid duplication of workload, as we do not implement the workload already existing in the OpenCL-SPIR-V translator when we develop the NVVM-SPIR-V translator. We show a diagram of the handling instructions in NVVM IR in Fig. \ref{fig:translator_diagram}. After translating, we can get SPIR-V IR shown in Fig. \ref{fig:different_IR}(b).

\subsection{Translate SPIR-V to OpenCL IR}
We need this phase for two reasons: 1) we want to execute SPIR-V on Vortex, and Vortex only accept the OpenCL IR as input; 2) SPIR-V is not human-readable; in other words, it's hard to debug. Thus, we add this phase to translate SPIR-V to OpenCL IR.
In this phase, we directly invoke the LLVM-SPIRV translator, described in Sec. \ref{sec:LLVM-SPIRV}.
After this step, we will get the OpenCL IR shown in Fig. \ref{fig:different_IR}(c).

\subsection{Execute OpenCL IR with Vortex}
The final step is to execute OpenCL IR with Vortex. For this phase, we have a customized  host code (host code is running on x86 while Kernel programs are running on RISC-V or extended RISC-V) to prepare for input data, set the arguments' types for the kernel function, allocate the memory buffer for the input and output, invoke Vortex's corresponding kernel launch function, and finally get the results and verify them. 

\section{Detailed Description and Experiments\label{sec:experiment}}
In this section, we describe the detailed information about what we need to do to support executing several CUDA features.

\subsection{Support built-in functions}

\subsubsection{Grid/Block information}
One of the most important features for CUDA is that it is SIMD.The key to implementing SIMD is to assign a different index for multiple threads, using a unique API. Although we have described how to support getting $threadIdx.x$ in CUDA on RISC-V, some detailed information still needs to explicitly be discussed.
NVVM will invoke different functions (Code. \ref{code:nvptx_index}) to get the index for different dimensions.

\begin{lstlisting}[caption={NVVM built-in function for getting index},label={code:nvptx_index},language=C]
; get block index, from x-z
@llvm.nvvm.read.ptx.sreg.ctaid.x()
@llvm.nvvm.read.ptx.sreg.ctaid.y()
@llvm.nvvm.read.ptx.sreg.ctaid.z()
; get thread index, from x-z
@llvm.nvvm.read.ptx.sreg.tid.x()
@llvm.nvvm.read.ptx.sreg.tid.y()
@llvm.nvvm.read.ptx.sreg.tid.z()
\end{lstlisting}

While SPIR-V will first load a global variable from memory and then extract different position for different dimensions, as shown in Code. \ref{code:opencl_index}. 
\begin{lstlisting}[caption={OpenCL built-in function for getting index},label={code:opencl_index},language=C]
; get block index, from x-z
variable = load BuiltInWorkgroupId;
extract global_variable, 0;
extract global_variable, 1;
extract global_variable, 2;
; get thread index, from x-z
variable = load BuiltInLocalInvocationId;
extract global_variable, 0;
extract global_variable, 1;
extract global_variable, 2;
\end{lstlisting}

To translate from an NVVM built-in function to SPIR-V, we have to replace the $@llvm.nvvm.read.ptx.sreg.ctaid.x()$'s call instruction with two consecutive instructions: load instruction and extract instruction. Besides, we have to analyze the function name of the call instruction to get the dimension the instruction needs and forward this dimension as an argument for the extract instruction. If this function name ends with $x$, it is for dimension 0. While function name ends with $y$ or $z$ for dimension 1 or 2.

\subsubsection{barrier}
NVVM has a simple API (Code. \ref{code:nvptx_sync}) to implement synchronization among all threads within a block.
\begin{lstlisting}[caption={NVVM built-in function for synchronization},label={code:nvptx_sync},language=C]
call void @llvm.nvvm.barrier0()
\end{lstlisting}
OpenCL's synchronization function, $barrier$, has a different prototype; it requests an argument for the memory address space needed to synchronize: $CLK\_LOCAL\_MEM\_FENCE$ for local memory and $CLK\_GLOBAL\_MEM\_FENCE$ for global memory. In fact, OpenCL's $barrier$ is a function to ensure the correct ordering of memory operations to global/local memory, not an actual synchronization barrier as $barrier0$ in NVVM. For NVVM, it $barrier0$ is used to synchronize all threads within a block to a same line. As the threads within a block can visit both global memory and local memory (share memory), we have to make sure they have the same order for these two memories. Thus, we translate the NVVM barrier to OpenCL's $barrier$ with a parameter to ensure the ordering for both local and global memory, as shown in Code. \ref{code:opencl_sync}.

\begin{lstlisting}[caption={OpenCL built-in function for synchronization},label={code:opencl_sync},language=C]
call void @barrier(i32
    CLK_LOCAL_MEM_FENCE|CLK_GLOBAL_MEM_FENCE)
\end{lstlisting}

\subsubsection{Atomic}
Several atomic operations are provided by CUDA, suck as $atomicAdd$, $atomicSub, atomicExch$, and so on. In NVVM, all these operations will be presented by the $atomicrmw$ instruction with different operations for add, sub, exchange, etc. (Code. \ref{code:nvptx_atomic}).
\begin{lstlisting}[caption={NVVM atomic operations},label={code:nvptx_atomic},language=C]
;atomicAdd(&data[0], 1);
%0 = atomicrmw add i32* %data, i32 1 seq_cst
;atomicSub(&data[0], -1);
%1 = atomicrmw add i32* %data, i32 -1 seq_cst
;atomicExch(&data[0], 1);
%2 = atomicrmw xchg i32* %data, i32 1 seq_cst
\end{lstlisting}

In OpenCL, these atomic operations are regarded as normal function calls (Code. \ref{code:opencl_atomic}). To translate from NVVM to OpenCL IR, our pipeline needs to extract the operation in each $atomicrmw$ instruction, and use this operation to choose the corresponding OpenCL's function name. After that, it directly copies the parameters in NVVM's instruction to OpenCL's instruction, as they have the same argument prototype (first argument: pointer, second argument: int32).

\begin{lstlisting}[caption={OpenCL atomic functions},label={code:opencl_atomic},language=C]
%0 = call i32 @atomic_add(i32* %data, i32 1)
%1 = call i32 @atomic_add(i32* %data, i32 -1)
%2 = call i32 @atomic_xchg(i32* %data, i32 1)
\end{lstlisting}

\subsection{Benchmark Experiments}
We also try to translate benchmarks. In Table. \ref{table:benchmark_test}, we record the translating results for applications in Rodinia\cite{che2009rodinia}. After we support features for grid/block information, barrier, and atomic instructions, we can succeed in translating most applications. However, there are still some applications we have not yet supported. These applications use either texture or some mathematical functions.
\begin{table}[]
\begin{tabular}{|c|c|c|}
\hline
application    & feature         & support? \\ \hline
b+tree         & -               & yes      \\ \hline
bfs            & -               & yes      \\ \hline
cfd            & double3 type    & yes      \\ \hline
huffman        & atomic          & yes      \\ \hline
pathfinder     & memory hierachy & yes      \\ \hline
gaussian       & -               & yes      \\ \hline
hotspot        & -               & yes      \\ \hline
hotspot3D      & -               & yes      \\ \hline
lud            & memory hierachy & yes      \\ \hline
nw             & -               & yes      \\ \hline
streamcluster  & -               & yes      \\ \hline
particlefilter & d2i             & on going       \\ \hline
backprop       & \_\_log2f       & on going       \\ \hline
lavaMD         & d2i             & on going       \\ \hline
kmeans         & texture         & no       \\ \hline
hybrid sort    & texture         & no       \\ \hline
leukocyte      & texture         & no       \\ \hline

\end{tabular}
\caption{Translating applications in Rodinia benchmark (Version: Vortex(v0.2.2) NVPTX-SPIR-V translator(v0.1.0)}
\label{table:benchmark_test}
\end{table}

\section{Conclusion\label{sec:conclusion}}
We have demonstrated a way to execute CUDA source code on an RISC-V back-end devices. To validate the feasibility, we build a pipeline that can succeed in executing multiple CUDA source codes with multiple features, including multi-thread,multi-block, atomic, and synchronization. Our pipeline comprises four steps: compiles CUDA source code into NVVM, translates NVVM and SPIR-V, uses modified POCL to emit object file, and finally, executes the generated object file on an open-source RISC-V GPU architecture. Except for the CUDA toolkit, which is required to compile NVVM, all other components are open-source and can be easily found in Github.

We also build a translator that supports translating NVVM into SPIR-V. This translator is lightweight and only relies on LLVM. It can be executed without the CUDA toolkit and GPUs. Our experiment results show that our translator can support most applications in Rodinia. In the future, we will try to support the remaining applications. In detail, we will support texture memory and mathematical functions, not only to convert from NVVM to SPIR-V, but also to support these corresponding libraries which will be needed when executing our generated SPIR-V on RISC-V GPUs.


\bibliographystyle{ACM-Reference-Format}
\nocite{*}
\bibliography{sample-base}

\begin{thebibliography}{15}


\ifx \showCODEN    \undefined \def \showCODEN     #1{\unskip}     \fi
\ifx \showDOI      \undefined \def \showDOI       #1{#1}\fi
\ifx \showISBNx    \undefined \def \showISBNx     #1{\unskip}     \fi
\ifx \showISBNxiii \undefined \def \showISBNxiii  #1{\unskip}     \fi
\ifx \showISSN     \undefined \def \showISSN      #1{\unskip}     \fi
\ifx \showLCCN     \undefined \def \showLCCN      #1{\unskip}     \fi
\ifx \shownote     \undefined \def \shownote      #1{#1}          \fi
\ifx \showarticletitle \undefined \def \showarticletitle #1{#1}   \fi
\ifx \showURL      \undefined \def \showURL       {\relax}        \fi
\providecommand\bibfield[2]{#2}
\providecommand\bibinfo[2]{#2}
\providecommand\natexlab[1]{#1}
\providecommand\showeprint[2][]{arXiv:#2}

\bibitem[\protect\citeauthoryear{Babej and J{\"a}{\"a}skel{\"a}inen}{Babej and
  J{\"a}{\"a}skel{\"a}inen}{2020}]%
        {babej2020hipcl}
\bibfield{author}{\bibinfo{person}{Michal Babej} {and} \bibinfo{person}{Pekka
  J{\"a}{\"a}skel{\"a}inen}.} \bibinfo{year}{2020}\natexlab{}.
\newblock \showarticletitle{HIPCL: Tool for Porting CUDA Applications to
  Advanced OpenCL Platforms Through HIP}. In
  \bibinfo{booktitle}{\emph{Proceedings of the International Workshop on
  OpenCL}}. \bibinfo{pages}{1--3}.
\newblock


\bibitem[\protect\citeauthoryear{Che, Boyer, Meng, Tarjan, Sheaffer, Lee, and
  Skadron}{Che et~al\mbox{.}}{2009}]%
        {che2009rodinia}
\bibfield{author}{\bibinfo{person}{Shuai Che}, \bibinfo{person}{Michael Boyer},
  \bibinfo{person}{Jiayuan Meng}, \bibinfo{person}{David Tarjan},
  \bibinfo{person}{Jeremy~W Sheaffer}, \bibinfo{person}{Sang-Ha Lee}, {and}
  \bibinfo{person}{Kevin Skadron}.} \bibinfo{year}{2009}\natexlab{}.
\newblock \showarticletitle{Rodinia: A benchmark suite for heterogeneous
  computing}. In \bibinfo{booktitle}{\emph{2009 IEEE international symposium on
  workload characterization (IISWC)}}. Ieee, \bibinfo{pages}{44--54}.
\newblock


\bibitem[\protect\citeauthoryear{Elsabbagh, Tine, Roshan, Lyons, Kim, Shim,
  Zhu, Lim, et~al\mbox{.}}{Elsabbagh et~al\mbox{.}}{2020}]%
        {elsabbagh2020vortex}
\bibfield{author}{\bibinfo{person}{Fares Elsabbagh}, \bibinfo{person}{Blaise
  Tine}, \bibinfo{person}{Priyadarshini Roshan}, \bibinfo{person}{Ethan Lyons},
  \bibinfo{person}{Euna Kim}, \bibinfo{person}{Da~Eun Shim},
  \bibinfo{person}{Lingjun Zhu}, \bibinfo{person}{Sung~Kyu Lim},
  {et~al\mbox{.}}} \bibinfo{year}{2020}\natexlab{}.
\newblock \showarticletitle{Vortex: OpenCL Compatible RISC-V GPGPU}.
\newblock \bibinfo{journal}{\emph{arXiv preprint arXiv:2002.12151}}
  (\bibinfo{year}{2020}).
\newblock


\bibitem[\protect\citeauthoryear{Grover and Lin}{Grover and Lin}{2012}]%
        {grover2012compiling}
\bibfield{author}{\bibinfo{person}{Vinod Grover} {and} \bibinfo{person}{Yuan
  Lin}.} \bibinfo{year}{2012}\natexlab{}.
\newblock \showarticletitle{Compiling CUDA and other languages for GPUs}. In
  \bibinfo{booktitle}{\emph{GPU Technology Conference (GTC)}}.
\newblock


\bibitem[\protect\citeauthoryear{J{\"a}{\"a}skel{\"a}inen, de~La~Lama,
  Schnetter, Raiskila, Takala, and Berg}{J{\"a}{\"a}skel{\"a}inen
  et~al\mbox{.}}{2015}]%
        {jaaskelainen2015pocl}
\bibfield{author}{\bibinfo{person}{Pekka J{\"a}{\"a}skel{\"a}inen},
  \bibinfo{person}{Carlos~S{\'a}nchez de La~Lama}, \bibinfo{person}{Erik
  Schnetter}, \bibinfo{person}{Kalle Raiskila}, \bibinfo{person}{Jarmo Takala},
  {and} \bibinfo{person}{Heikki Berg}.} \bibinfo{year}{2015}\natexlab{}.
\newblock \showarticletitle{pocl: A performance-portable OpenCL
  implementation}.
\newblock \bibinfo{journal}{\emph{International Journal of Parallel
  Programming}} \bibinfo{volume}{43}, \bibinfo{number}{5}
  (\bibinfo{year}{2015}), \bibinfo{pages}{752--785}.
\newblock


\bibitem[\protect\citeauthoryear{Keryell, Reyes, and Howes}{Keryell
  et~al\mbox{.}}{2015}]%
        {keryell2015khronos}
\bibfield{author}{\bibinfo{person}{Ronan Keryell}, \bibinfo{person}{Ruyman
  Reyes}, {and} \bibinfo{person}{Lee Howes}.} \bibinfo{year}{2015}\natexlab{}.
\newblock \showarticletitle{Khronos SYCL for OpenCL: a tutorial}. In
  \bibinfo{booktitle}{\emph{Proceedings of the 3rd International Workshop on
  OpenCL}}. \bibinfo{pages}{1--1}.
\newblock


\bibitem[\protect\citeauthoryear{Kessenich, Ouriel, and Krisch}{Kessenich
  et~al\mbox{.}}{2018}]%
        {kessenich2018spir}
\bibfield{author}{\bibinfo{person}{John Kessenich}, \bibinfo{person}{Boaz
  Ouriel}, {and} \bibinfo{person}{Raun Krisch}.}
  \bibinfo{year}{2018}\natexlab{}.
\newblock \showarticletitle{SPIR-V Specification}.
\newblock \bibinfo{journal}{\emph{Khronos Group}}  \bibinfo{volume}{3}
  (\bibinfo{year}{2018}).
\newblock


\bibitem[\protect\citeauthoryear{Kontogiannis, Martin, Wong, Gregory,
  M{\"u}ller, and Mylopoulos}{Kontogiannis et~al\mbox{.}}{2010}]%
        {kontogiannis2010code}
\bibfield{author}{\bibinfo{person}{Kostas Kontogiannis},
  \bibinfo{person}{Johannes Martin}, \bibinfo{person}{Kenny Wong},
  \bibinfo{person}{Richard Gregory}, \bibinfo{person}{Hausi M{\"u}ller}, {and}
  \bibinfo{person}{John Mylopoulos}.} \bibinfo{year}{2010}\natexlab{}.
\newblock \showarticletitle{Code migration through transformations: An
  experience report}.
\newblock In \bibinfo{booktitle}{\emph{CASCON First Decade High Impact
  Papers}}. \bibinfo{pages}{201--213}.
\newblock


\bibitem[\protect\citeauthoryear{Krizhevsky, Sutskever, and Hinton}{Krizhevsky
  et~al\mbox{.}}{2012}]%
        {krizhevsky2012imagenet}
\bibfield{author}{\bibinfo{person}{Alex Krizhevsky}, \bibinfo{person}{Ilya
  Sutskever}, {and} \bibinfo{person}{Geoffrey~E Hinton}.}
  \bibinfo{year}{2012}\natexlab{}.
\newblock \showarticletitle{Imagenet classification with deep convolutional
  neural networks}.
\newblock \bibinfo{journal}{\emph{Advances in neural information processing
  systems}}  \bibinfo{volume}{25} (\bibinfo{year}{2012}),
  \bibinfo{pages}{1097--1105}.
\newblock


\bibitem[\protect\citeauthoryear{Kuznetsov and Stegailov}{Kuznetsov and
  Stegailov}{2019}]%
        {kuznetsov2019porting}
\bibfield{author}{\bibinfo{person}{Evgeny Kuznetsov} {and}
  \bibinfo{person}{Vladimir Stegailov}.} \bibinfo{year}{2019}\natexlab{}.
\newblock \showarticletitle{Porting CUDA-Based Molecular Dynamics Algorithms to
  AMD ROCm Platform Using HIP Framework: Performance Analysis}. In
  \bibinfo{booktitle}{\emph{Russian Supercomputing Days}}. Springer,
  \bibinfo{pages}{121--130}.
\newblock


\bibitem[\protect\citeauthoryear{Lattner and Adve}{Lattner and Adve}{2004}]%
        {lattner2004llvm}
\bibfield{author}{\bibinfo{person}{Chris Lattner} {and} \bibinfo{person}{Vikram
  Adve}.} \bibinfo{year}{2004}\natexlab{}.
\newblock \showarticletitle{LLVM: A compilation framework for lifelong program
  analysis \& transformation}. In \bibinfo{booktitle}{\emph{International
  Symposium on Code Generation and Optimization, 2004. CGO 2004.}} IEEE,
  \bibinfo{pages}{75--86}.
\newblock


\bibitem[\protect\citeauthoryear{Munshi}{Munshi}{2009}]%
        {munshi2009opencl}
\bibfield{author}{\bibinfo{person}{Aaftab Munshi}.}
  \bibinfo{year}{2009}\natexlab{}.
\newblock \showarticletitle{The opencl specification}. In
  \bibinfo{booktitle}{\emph{2009 IEEE Hot Chips 21 Symposium (HCS)}}. IEEE,
  \bibinfo{pages}{1--314}.
\newblock


\bibitem[\protect\citeauthoryear{Nguyen, Nguyen, and Nguyen}{Nguyen
  et~al\mbox{.}}{2016}]%
        {nguyen2016mapping}
\bibfield{author}{\bibinfo{person}{Trong~Duc Nguyen}, \bibinfo{person}{Anh~Tuan
  Nguyen}, {and} \bibinfo{person}{Tien~N Nguyen}.}
  \bibinfo{year}{2016}\natexlab{}.
\newblock \showarticletitle{Mapping API elements for code migration with vector
  representations}. In \bibinfo{booktitle}{\emph{2016 IEEE/ACM 38th
  International Conference on Software Engineering Companion (ICSE-C)}}. IEEE,
  \bibinfo{pages}{756--758}.
\newblock


\bibitem[\protect\citeauthoryear{Perkins}{Perkins}{2017}]%
        {perkins2017cuda}
\bibfield{author}{\bibinfo{person}{Hugh Perkins}.}
  \bibinfo{year}{2017}\natexlab{}.
\newblock \showarticletitle{CUDA-on-CL: a compiler and runtime for running
  NVIDIA{\textregistered} CUDA™ C++ 11 applications on OpenCL™ 1.2
  Devices}. In \bibinfo{booktitle}{\emph{Proceedings of the 5th International
  Workshop on OpenCL}}. \bibinfo{pages}{1--4}.
\newblock


\bibitem[\protect\citeauthoryear{Silveira, Avila, Barreto, and Navaux}{Silveira
  et~al\mbox{.}}{2000}]%
        {silveira2000dpc++}
\bibfield{author}{\bibinfo{person}{Andr{\'e} Silveira},
  \bibinfo{person}{Rafael~Bohrer Avila}, \bibinfo{person}{Marcos~E Barreto},
  {and} \bibinfo{person}{Philippe Olivier~Alexandre Navaux}.}
  \bibinfo{year}{2000}\natexlab{}.
\newblock \showarticletitle{DPC++: Object-Oriented Programming Applied to
  Cluster Computing.}. In \bibinfo{booktitle}{\emph{PDPTA}}.
\newblock


\end{thebibliography}
\end{document}